\documentclass[aps,pra,nobibnotes,twocolumn,superscriptaddress]{revtex4-1}
\usepackage{graphicx}
\usepackage{mathrsfs}
\usepackage{bm}
\usepackage{amsmath}
\usepackage{dcolumn}
\usepackage{epstopdf}
\usepackage{dsfont}
\usepackage{amssymb}
\usepackage{tabularx}
\usepackage{array}
\usepackage{float}
\usepackage{color}
\usepackage{epstopdf}
\usepackage{mathrsfs}
\usepackage[colorlinks, linkcolor=blue,anchorcolor=blue,citecolor=blue,urlcolor=blue]{hyperref}

\begin{document}

\preprint{APS/123-QED}

\title{Hardcore bosonic domain walls on honeycomb lattice}
\author{Xingchuan Zhu}
\affiliation{Department of Physics, Beijing Normal University, Beijing, 100875, China}

\author{Bo Li}
\affiliation{School of Computer Science and Engineering, Beihang University, Beijing, 100191, China}

\author{Huaiming Guo}
\email{hmguo@buaa.edu.cn}
\affiliation{Department of Physics, Key Laboratory of Micro-Nano Measurement-Manipulation and Physics (Ministry of Education), Beihang University,
Beijing, 100191, China}

\author{Shiping Feng}
\affiliation{Department of Physics, Beijing Normal University, Beijing, 100875, China}


\begin{abstract}
Linelike hardcore bosonic domain walls in a staggered potential on honeycomb lattice are studied using quantum Monte Carlo simulations. The phase diagrams of ribbons with zigzag and armchair domain walls are mapped, which contain superfluid and insulator phases at various fillings. In the $\rho=\frac{1}{2}$ insulator, the domain wall separates two charge-density-wave (CDW) regions with opposite Berry curvatures. Associated with the change of topological properties, superfluid transport occurs down the domain wall. The superfluid density associated with a zigzag domain wall is much larger than that of an armchair domain wall due to the different arrangements of occupied and unoccupied sites along the domain wall. Our results provide a concrete context to study bosonic topological phenomena, which may be simulated experimentally using bosonic cold atoms trapped in optical lattices.
\end{abstract}

\maketitle

%\tableofcontents
%\textit{Introduction.-}
\section{Introduction}

The rise of topological insulator has made the study of new topological phases one of the most active fields in condensed matter physics\cite{hasan2010,qixl2011,franz2013,bansil2016,ashvin2018}. In the many studies, an important direction is to generalize the many known topological properties to bosonic systems\cite{senthil2013,frank2015}. However since bosons tend to condensate and the band structures collapse, such generalizations usually are not direct\cite{varney2010,hofstetter2015}.

The Su-Schrieffer-Heeger model is the simplest one-dimensional lattice with nontrivial topology\cite{su1979,Lienhard2019}. Periers instability distorts the lattice, and the hopping integrals become dimerized. There are two ways of choosing the unit cell: strong or weak intra-unit-cell hopping, which corresponds to topological or trivial insulator. A domain wall between the two ground states is a topological object, and gives rise to zero-energy midgap states. In the low-energy continuum theory, the zero mode can be explained in terms of Jackiw-Rebbi model\cite{jackiw1976}, i.e., a Dirac equation with a kink in the spatial-variant mass.

The topological domain walls have been intensively studied in graphene because of their fascinating physical properties. Linelike domain walls can be created in the mass pattern in graphene gapped by a staggered on-site potential\cite{semenoff2008,yang2019}. They support midgap states, which are localized in the vicinity of the domain wall and propagate along the length.
Topological one-dimensional domain walls can also be formed in bilayer graphene, which result from either electrostatic lateral confinement\cite{martin2008}, or the transition between AB- and BA- stacking orders\cite{zhang2013,eun-ahkim2013}. Such bilayer domain walls feature one-dimensional valley-polarized conducting channels, and have been observed experimentally\cite{ju2015,helin2016,li2016,jiang2018}.

An interesting question is whether such topological kink states associated with domain wall can be generalized from fermions to bosons, realizing superfluid propagating down the bosonic domain wall. In the paper, we study hardcore bosonic versions of domain walls in honeycomb lattice gapped by a staggered potential. Using quantum Monte Carlo simulations, the phase diagrams of ribbons with zigzag and armchair domain walls are mapped, which contain superfluid and insulator phases at various fillings. Specifically the $\rho=\frac{1}{2}$ insulator is a domain-wall phase, where the domain wall separates two CDW regions with opposite Berry curvatures. Associated with the change of topological properties, superfluid transport occurs down the domain wall. The superfluid density associated with a zigzag domain wall is much larger than that of an armchair domain wall due to the different arrangements of occupied and unoccupied sites along the domain wall. Recently, it is proposed that the zigzag domain wall can be created by nearest-neighbor repulsion in a self-organized way\cite{zhu2019}. These results are also experimentally related to bosonic cold atoms trapped in optical lattices.

This paper is organized as follows: Section 2 introduces the precise model we will investigate, along with our computational methodology. Section 3 presents the phase diagram of zigzag domain wall from numerical calculations. Section 4 discusses the topological property of the zigzag domain wall. Section 5 shows the results of armchair domain wall, and is followed by some further discussion and interpretation in Sec.6. One appendix addressing the eigenenergies of the zigzag domain wall at $k_x=\pi$ is also included.

%textit{The model and method.-}
\section{The model and method}

\begin{figure}[htbp]
\centering \includegraphics[width=9.cm]{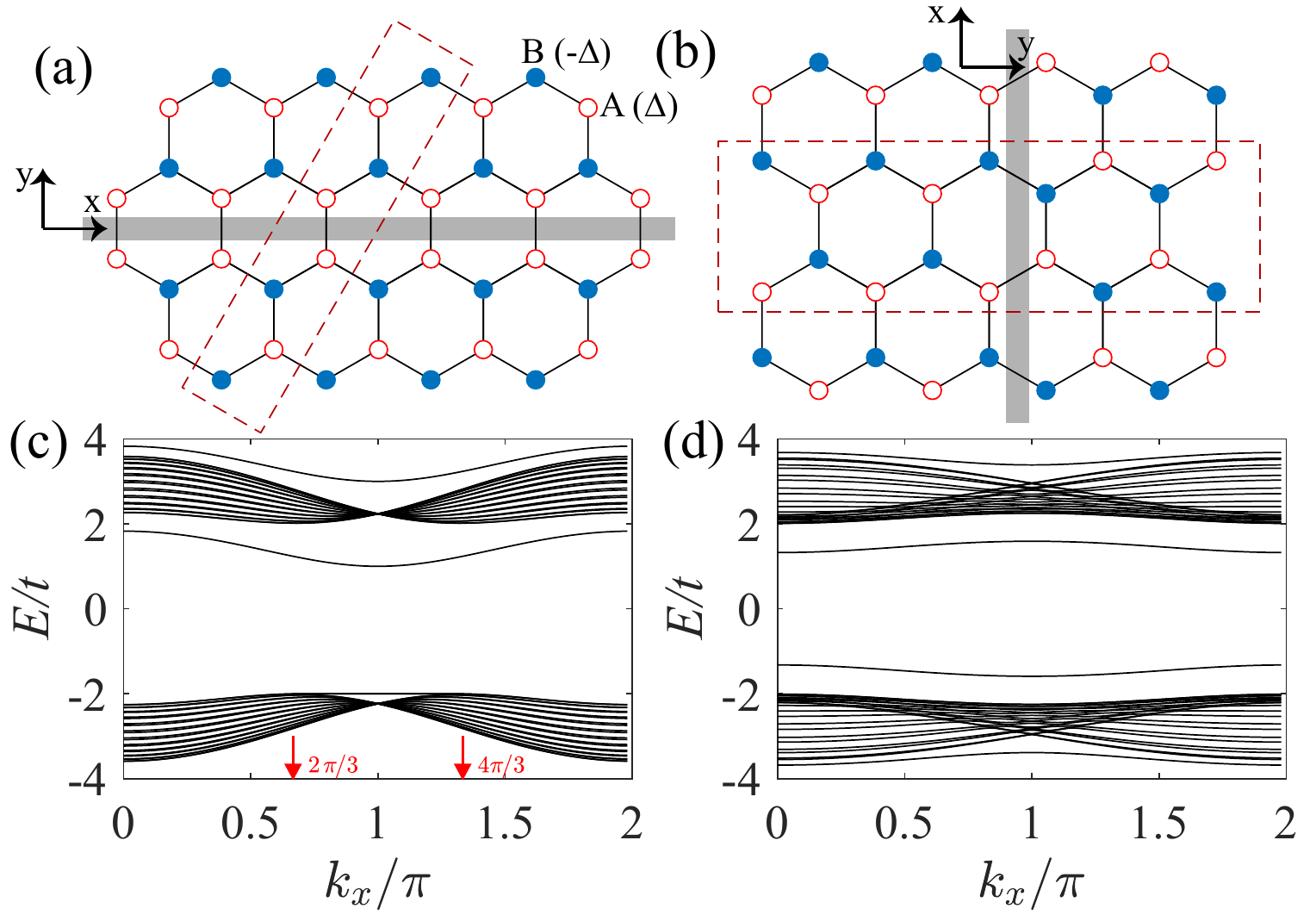} \caption{Schematic illustrations of (a) a zigzag domain wall on a $W=4$ ribbon and (b) a armchair domain wall on a $W=5$ ribbon. The positions of the domain walls are marked by thick solid lines. The ribbons are translational invariant along $x$-direction, and the unit cells are marked in dashed rectangles.  The band structures of honeycomb ribbons with (c) a zigzag domain wall and (d) an armchair domain wall in the middle. The widths of the ribbons in (c) and (d) are $W=24$ and $12$, respectively.  The strength of the staggered potential is $\Delta=2t$.}
\label{fig1}
\end{figure}

We consider a tight-binding model on honeycomb lattice described by the following hardcore Bose-Hubbard model\cite{bloch2008,Hoang2010,richard2007,richard2007a}
\begin{eqnarray}\label{eq1}
H&=&-t\sum_{\langle i,j\rangle} (b_i^{\dagger}b_{j}+\text{H.c.})+\sum_i\Delta_i n_i-\mu \sum_i n_i,
\end{eqnarray}
where $b_i$ ($b_i^{\dagger}$) is the hardcore bosonic annihilation (creation) operator, $n_i=b_i^{\dagger}b_i$ is the number operator of bosons. The occupying number of hardcore bosons is $0$ or $1$ on each site. Hence, the hardcore bosons obey commutation relation $[b_i,b_j^{\dagger}]=0$ for sites $i\neq j$ but anticommutation relation $\{b_i,b_i^{\dagger}\}=1$ for a single site $i$. This hardcore condition corresponds to the limit of infinite on-site interactions, thus the model is a strongly interacting one.  The first term in Eq.~(\ref{eq1}) is the nearest-neighbor (NN) hopping, and the hopping amplitude $t$ will be taken as the unit of energy $(t=1)$ in our calculation. The second term is an on-site potential with $\Delta_i$ describing the pattern of the on-site energy. The last term denotes the chemical potential, which controls the average density of the system.

The honeycomb lattice has two points in the
unit cell, so that it belongs to the class of Bravais lattices with a
basis. And that this two basis points are those that define the two
triangular sublattices, which we denote A and B sublattices in Fig.1(a) and (b). Performing a particle-hole transformation $b_i^{\dagger}(b_i)\rightarrow h_{i}(h_{i}^{\dagger})$,
the model in Eq.~(\ref{eq1}) becomes
 \begin{eqnarray}\label{eq2}
H_{h}&=&-t\sum_{\langle i,j\rangle} (h_i^{\dagger}h_{j}+\text{H.c.})- \sum_i \Delta_i n^h_i \\ \nonumber
&+&\mu \sum_i n^{h}_i+E_0,
\end{eqnarray}
where $n^h_i=h^{\dagger}_{i}h_{i}$ is the number operator of holes, and $E_{0}=-N\mu+\sum_{i}^{N}\Delta_i$ is a constant with $N$ the total number of sites. If the sign of $\Delta_i$ is irrelevant, the Hamiltonian at $\mu$ in the hole representation is equivalent to that at $-\mu$ in the particle representation. The Hamiltonian is symmetric about $\mu=0$, and so are the physical quantities and the phase diagram. This is the case for the armchair ribbon shown in Fig.\ref{fig1}(b), where the $\Delta_i$ and $-\Delta_i$ configurations are related to each other by a $\pi$ rotation (mirror transformation), under which the Hamiltonian is definitely invariant.

When the on-site energy represents a staggered potential, i.e., $\Delta_i=\Delta(-\Delta)$ for $A$($B$) sublattice, the \textit{bulk} energy spectrum of Eq.(1) has two branches,
 \begin{eqnarray}
E_{\bf k}=\pm\sqrt{t^2(2\cos\frac{\sqrt{3}}{2}k_x+\cos\frac{3}{2}k_y)^2+t^2\sin^2\frac{3}{2}k_y+\Delta^2},
&&
\end{eqnarray}
where ${\bf k}=(k_x, k_y)$ are momenta. The spectrum is symmetric about $E=0$, and has a gap with the size $2\Delta$.

In the paper, we focus on the patterns with domain walls in the middle of the geometries (see Fig.~1)\cite{semenoff2008}. Such domain walls break the regularity of the staggered on-site potential, and are indeed \textit{defects} in the pattern of the staggered potential discussed above. The system considered is only translation invariant
along the $x$-axis. By working with periodic boundary conditions
along this axis, the band structures can be numerically obtained. For a zigzag domain wall, two dispersive bands associated with the domain wall are obvious for large $\Delta$, which are separated from the bulk spectrum. Since the state at $k_x=\pi$ is localized in the vertical bonds of the domain wall, the eigenvalues can be analytically determined. The matrix of an isolated two-site bond is
\begin{eqnarray}\label{eq3}
H_{k_x=\pi}=\left(
              \begin{array}{cc}
                \Delta & -t \\
                -t & \Delta \\
              \end{array}
            \right).
\end{eqnarray}
Thus the eigenvalues at $k_x=\pi$ are $\Delta+t$ and $\Delta-t$, which corresponds to the bonding and antibonding states, respectively.
In the spectrum, there are also two flat bands connecting the Dirac points, which are due to the zigzag edges. At $k_x=\pi$, the state is totally localized on the outmost site, and the eigenvalue $-\Delta$ is directly obtained (see the Appendix). A domain wall can also be created along the armchair direction. There are four bands associated with the domain wall. They are easily understood from the case of large $\Delta$, and their values are determined by two dimers with uniform on-site energies $\pm \Delta$, which are $\pm t\pm\Delta$. Two of them ($\Delta-t, -\Delta+t$) are in the gap, and the other two ($\Delta+t, -\Delta-t$) are outside the bulk bands which are $\Delta, -\Delta$. We plot the band structure for $\Delta=2$ in Fig.\ref{fig1}. Although the four bands become dispersive, they are still well separated from the bulk bands\cite{bulk_bands}. The corresponding wavefunctions are mainly localized near the domain wall.

In the following discussions, we employ the approach of stochastic series expansion (SSE) quantum Monte Carlo (QMC) method~\cite{sandvik2002,syljuasen2003} with directed loop updates to study the model in Eq.(1). The SSE method expands the partition function in power series and the trace is written as a sum of diagonal matrix elements. The directed loop updates make the simulation very efficient~\cite{Bauer2011,fabien2005,pollet2004}. Our simulations are on finite lattices with the total number of sites $N=2\times W \times L$ with $W$ the width and $L$ the length of a ribbon. There are no approximations
causing systematic errors, and the discrete configuration space can be sampled without floating
point operations. The temperature is set to be low enough to obtain the ground-state properties. For such bosonic systems, the notorious sign problem in the QMC approach can be avoided.

\begin{figure}[htbp]
\centering \includegraphics[width=9.cm]{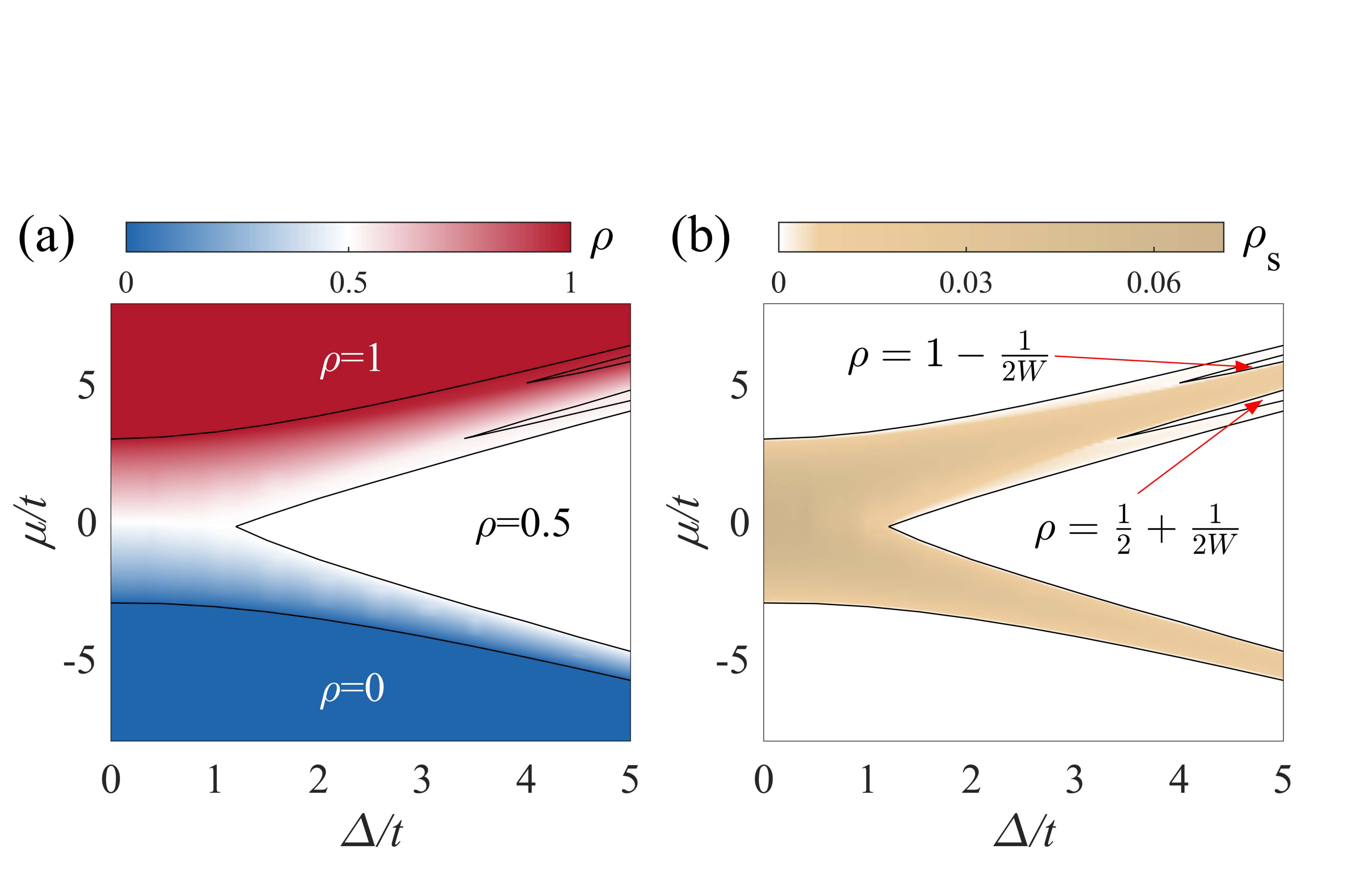} \caption{The phase diagram of the honeycomb lattice ribbons with a domain wall along the zigzag direction in the $(\Delta, \mu)$ plane, which contains superfluid and insulators at various fillings. The false color plots of the average density and the superfluid density are shown in (a) and (b), respectively.}
\label{fig2}
\end{figure}

\begin{figure}[htbp]
\centering \includegraphics[width=7.5cm]{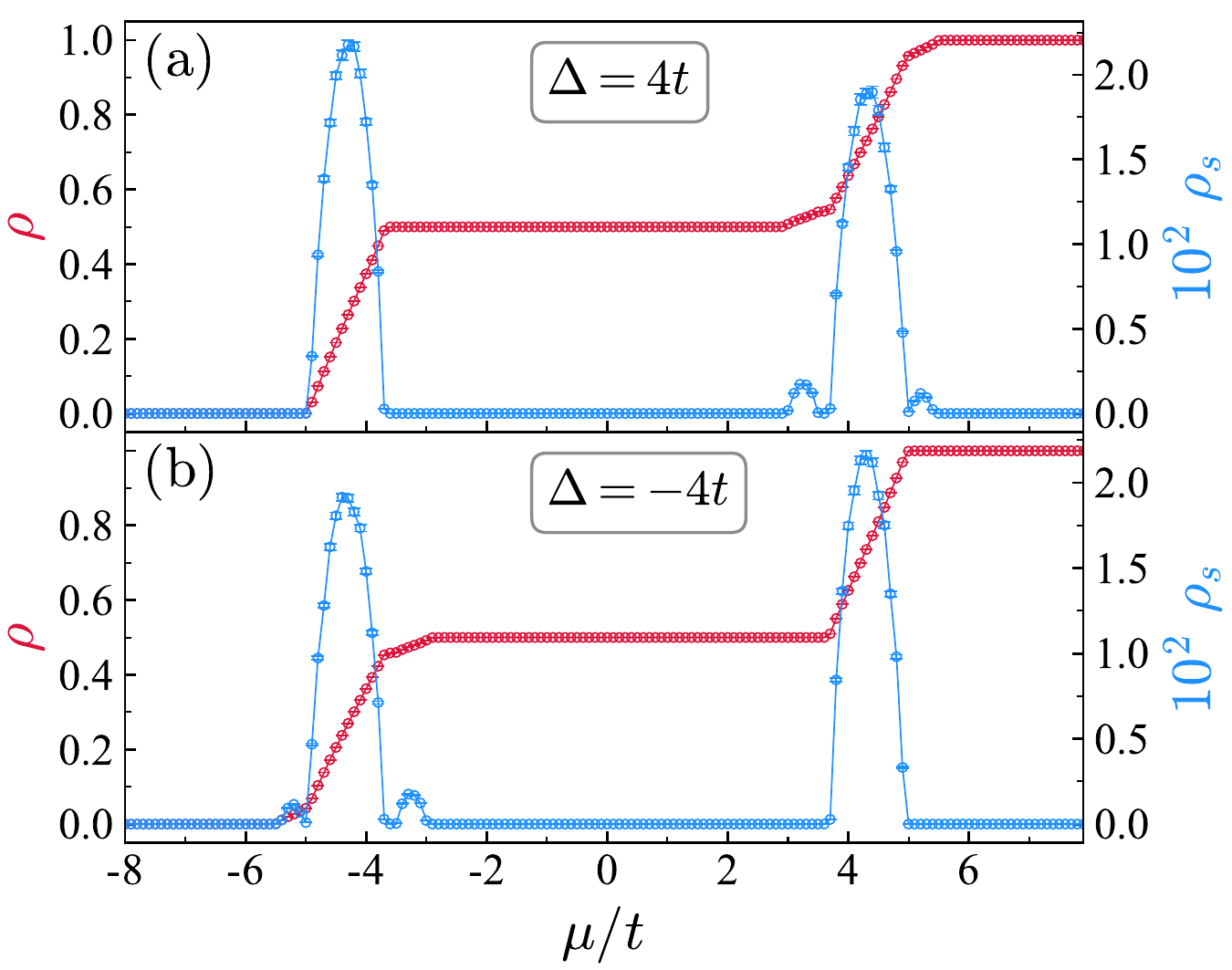} \caption{The average density and superfluid density as a function of $\mu$ at $\Delta=\pm 4t$ on a $W=12$ and $L=24$ ribbon. }
\label{fig3}
\end{figure}

\section{The phase diagram of the zigzag domain wall}
%\textit{The phase diagram.-}
The phase diagrams of the ribbons with the domain wall along the zigzag direction are shown in Fig.\ref{fig2}.
In the atomic limit ($t=0$), the system is a $\rho=\frac{1}{2}$ insulator for $-\Delta<\mu<\Delta$. For $\mu>\Delta$, all sites become occupied, and it is a $\rho=1$ Mott insulator.
In the presence of hoppings, one observes that the atomic insulators persist at large $\Delta$. As $\Delta$ is decreased, the range in the chemical potential also decreases, and completely disappear at a critical value $\Delta_c\sim t$. For large enough $\Delta>0$ ($\Delta<0$), there appear two small regions between the $\rho=\frac{1}{2}, 1$ ($\rho=0, \frac{1}{2}$) plateaus, which correspond to $\rho=\frac{1}{2}+\frac{1}{2W}, 1-\frac{1}{2W}$ ($\rho=\frac{1}{2W}, \frac{1}{2}-\frac{1}{2W}$) insulators, respectively. All the insulators are separated by incommensurate superfluid regions.

The various quantum phases are characterized by the average density $\rho=\frac{1}{N}\sum_{i}n_{i}$ and the superfluid density
$\rho_{s}=\frac{\langle W_x^{2}+W_y^2\rangle}{4\beta t}$,
where $W_{x(y)}$ is the winding number of the world line along $x(y)$-direction, and $\beta$ is the inverse temperature\cite{pollock1987}. An insulator is characterized by plateaus of $\rho$ with $\rho_s=0$, while a superfluid phase is characterized by a nonzero $\rho_s$.  In the phase diagram, the average density and superfluid density are plotted using false colors, and the features of different quantum phases are clearly demonstrated.
Specifically we plot $\rho$ and $\rho_s$ as a function of $\mu$ along two typical cuts $\Delta=\pm 4t$, on which all phases in the phase diagram are encountered, as shown in Fig.~\ref{fig3}. The average density exhibits a series of plateaus at commensurate fillings, on which the superfluid density vanishes. These plateaus correspond to the incompressible insulating phases, whose gaps are given by the widths of the plateaus. Between the insulators, the average density and the superfluid density are finite, and the system is in a superfluid phase.

Interestingly, the two small superfluid regions in each panel of Fig.~\ref{fig3} are associated with the existence of the domain wall. For $\Delta>0$, the sites connected by the vertical bonds on the domain wall are empty in the $\rho=\frac{1}{2}$ insulator. As the chemical potential is further increased, such sites are energetically favored for the added bosons to reside on, rather than those inside the CDW phase. It is because the bosons can hop within the vertical bonds without experiencing potential barriers, and have a large gain of kinetic energy, which is proportional to $t$. As shown in Fig.\ref{fig3}, the average density continuously increases from the $\rho=\frac{1}{2}$ plateaus, and the superfluid density becomes finite, implying the added bosons induce superfluid transport along the domain wall. When the domain wall is full, i.e., one boson in each vertical bond, the superfluid density becomes zero, and the system becomes a $\rho=\frac{1}{2}+\frac{1}{2W}$ insulator. Such a insulator is related to the occupancy of the sites on the domain wall. Its region tends to vanish in the $W\to \infty$ limit, as a result of the vanishing ratio of the number of the domain wall sites to the total number of sites.
The superfluid density is maximum about at the density when the domain wall is half filled, which results from the balance between the number of bosons and the free space.

\begin{figure}[htbp]
\centering \includegraphics[width=9cm]{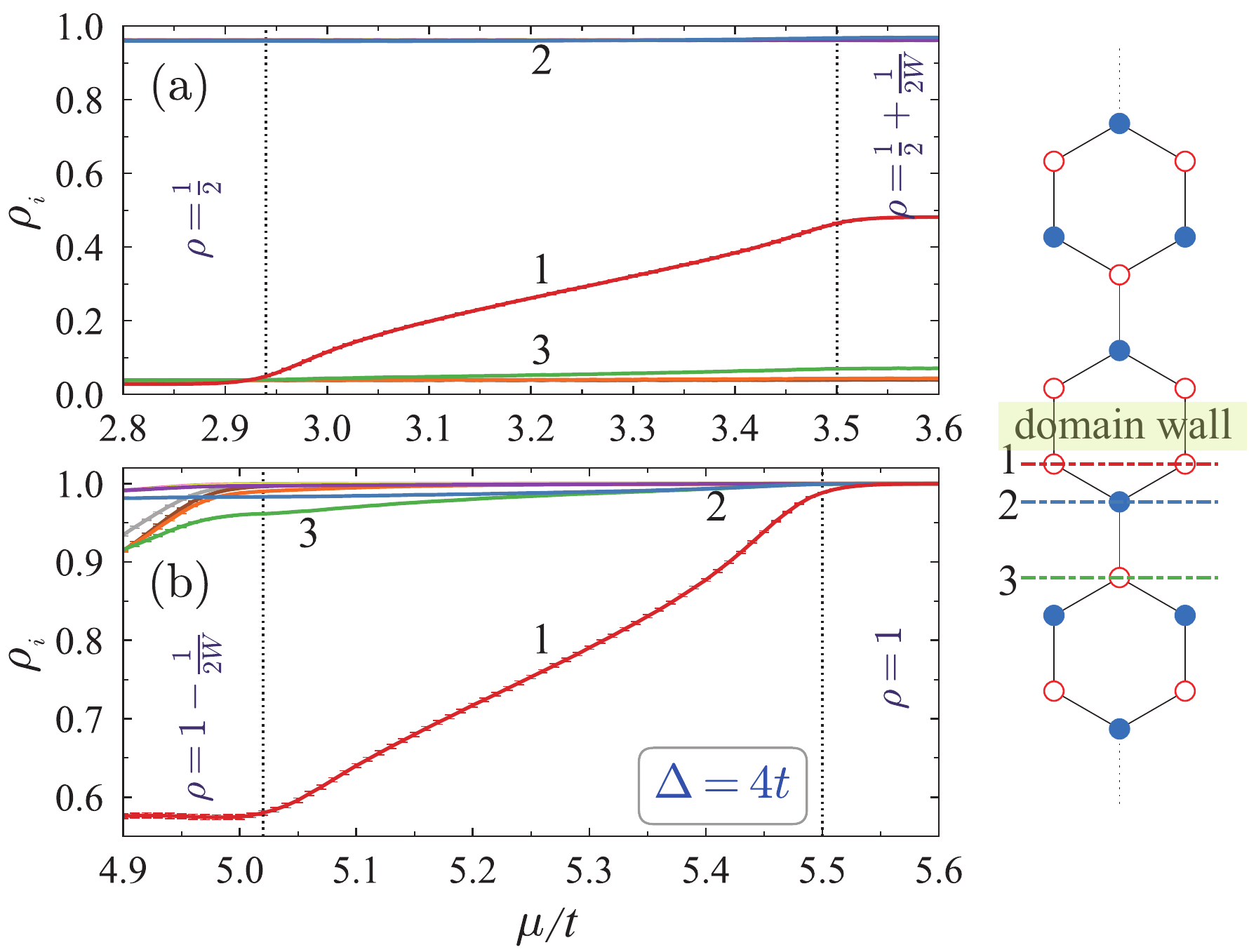} \caption{The local densities as a function of the chemical potential around the small superfluid regions: (a), the one between $\rho=\frac{1}{2}$ and $\rho=\frac{1}{2}+\frac{1}{2W}$ insulators; (b), that between $\rho=1-\frac{1}{2W}$ and $\rho=1$ insulators. Due to the mirror symmetry about the domain wall, we only consider inequivalent sites, which are marked in the right lattice. The strength of the staggered potential is $\Delta=4t$, and the lattice has a width $W=12$ and a length $L=24$.}
\label{fig4}
\end{figure}

To show the distribution of the added bosons directly, the local densities of inequivalent sites are plotted in Fig.4. In the $\rho=\frac{1}{2}$ insulator, the densities on the sites with the potential $|\Delta|(-|\Delta|)$ have small (large) values [see Fig.\ref{fig4}(a)], which are  coincident with the CDW order. Between the $\rho=\frac{1}{2}$ and $\frac{1}{2}+\frac{1}{2W}$ insulators, while the local densities of other sites are almost unchanged, those on the domain wall increase dramatically, indicating the added bosons enter the domain wall. In the $\rho=\frac{1}{2}+\frac{1}{2W}$ insulator, the local density on each site of the domain wall is about $\rho_{1}\sim 0.5$, implying there is one boson in each vertical bond of the domain wall. Since the sites connected by such bonds have the same potentials, the bosons can hop freely between them, forming dimers.
Such kind of dimers are also formed in the $\rho=1-\frac{1}{2W}$ insulator. As shown in the left part of Fig.\ref{fig4}(b) which corresponds to the above insulator, while all other sites are almost occupied, the sites on the domain wall are near half filling, implying there is approximately one boson in each vertical bond of the domain wall, thus dimers are formed. The small superfluid regions for $\Delta<0$ have similar physical origin.

\section{The topological property of the zigzag domain wall}
Due to bulk-boundary correspondence, the appearance of the domain-wall and edge states is a manifestation of the nontrivial bulk topological property. Next we study the topological property of the Bose-Hubbard model with a uniform staggered potential on a lattice with periodic boundary condition in both directions.
The model is equivalent to a $S=1/2$ $XXZ$ model through a mapping $S^{+}_i=b^{\dag}_i$ and $S^z_i=n_i-\frac{1}{2}$~\cite{Owerre2016,zhu2019}. Spin obeys commutation relations,
\begin{align}
\left[S_{\alpha,i},S_{\beta,j}\right]=i\hbar \varepsilon _{\alpha\beta\gamma}S_{
\gamma,i}\delta_{ij}
\end{align}
where $\varepsilon _{\alpha\beta\gamma}$ is the Levi-Civita symbol; $\alpha,\beta,\gamma\in (x,y,z)$ represent the spin direction; $i,j$ are the sites on which the spins are located. While we have the commutation relation $\left[S^{+}_{i},S^{-}_{i}\right]=2S^z_{i}$ for spin operators, there is also the anticommutation relation $\{S^{+}_{i},S^{-}_{i}\}=I$ analogous to that of hardcore boson.

\begin{figure}[htbp]
\centering \includegraphics[width=9.cm]{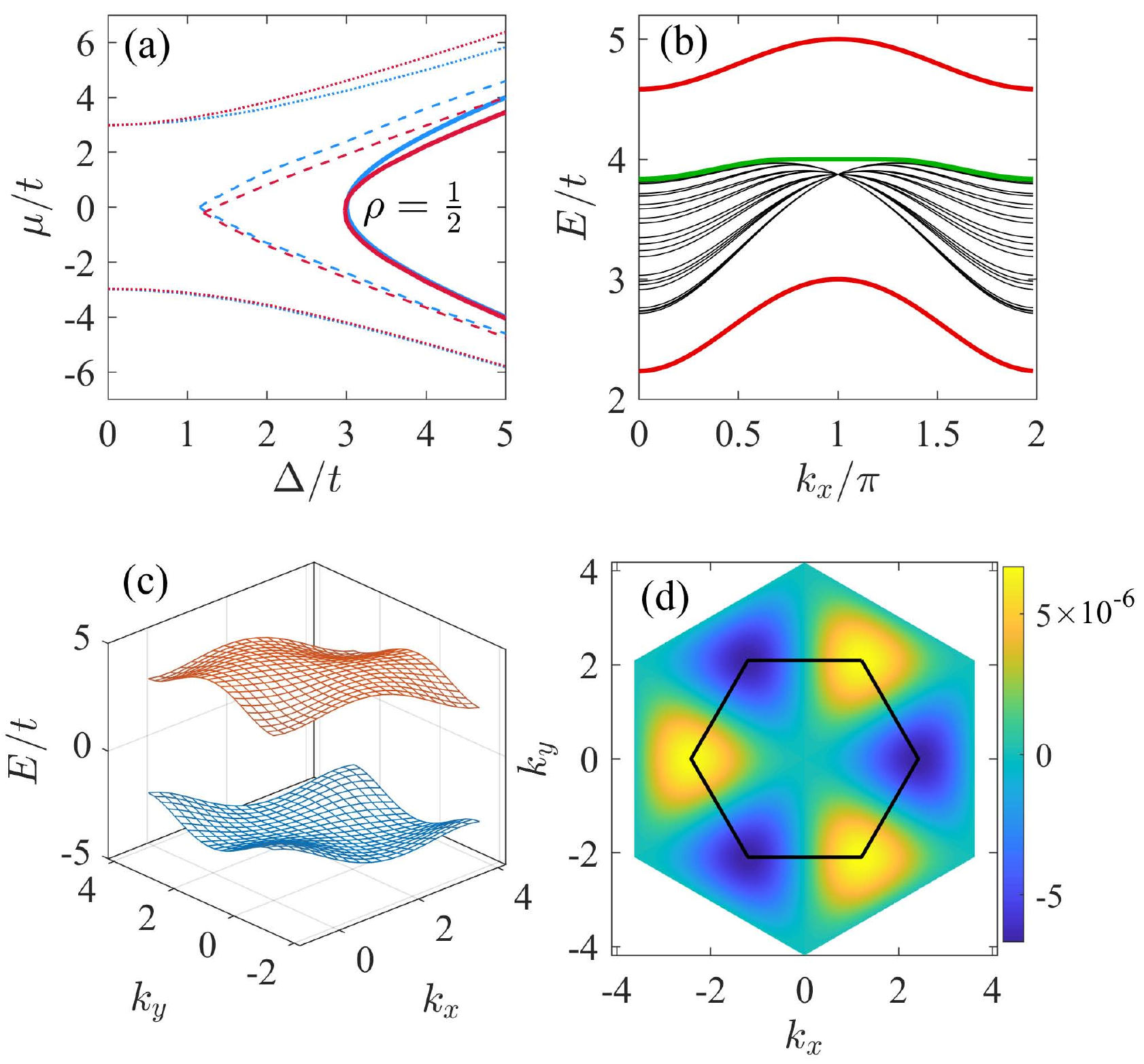}
\caption{(a) The phase diagrams of the Bose-Hubbard model on the periodic honeycomb lattice with a staggered potential (blue curves)~\cite{guo2016} and zigzag ribbon with a domain wall (red curve). The thick solid (thin dashed) curves are the phase boundaries from the spin-wave approximation (the QMC method). (b) The excitation spectrum on a $W=12$ ribbon with a zigzag domain wall in the middle. The red curves represent states localized near the domain wall. The green curves are two-fold degenerate, and are associated with the zigzag edges. (c) The magnon band structure, where $E^{+}, E^{-}$ are identical and we plot $-E^{-}$ to display it. (d) The Berry curvature associated with the
upper magnon band, which differs from that of the lower band by a sign. The first Brillouin zone is marked by black lines. The parameters are $\Delta=4t$.
}
\label{fig5}
\end{figure}

Using Holstein-Primakoff transformation and linear spin-wave approximation, the spin operators are expressed in term of bosonic creation and annihilation operators. The honeycomb lattice is a bipartite one. The transformation on sublattice $A$ ($\Delta<0$) is defined as
\begin{align}\label{eq5x}
S^+_{A,i}=a_{i,A}, S^-_{A,i}=a^{\dagger}_{i,A},S^z_{A,i}=\frac{1}{2}-a^{\dagger}_{i,A}a_{i,A}.
\end{align}
On sublattice $B$ ($\Delta>0$), the spin is in the opposite direction for the antiferromagnet order. Thus the spin operators are defined as
\begin{align}\label{eq6x}
S^+_{B,i}=a^{\dagger}_{i,B},S^-_{B,i}= a_{i,B},S^z_{B,i}=a^{\dagger}_{i,B} a_{i,B}-\frac{1}{2}.
\end{align}
Then the bosonic tight binding Hamiltonian becomes
\begin{eqnarray}\label{eq7}
H=&-&t\sum_{\langle i,j\rangle} (a_{i,A}a_{j,B}+a_{i,A}^{\dagger}a_{j,B}^{\dagger}) \\ \nonumber
&-&(\Delta+\mu)\sum_{i\in A}(1-a_{i,A}^{\dagger}a_{i,A})+(\Delta-\mu)\sum_{i\in B}a_{i,B}^{\dagger}a_{i,B}.
\end{eqnarray}
Ignoring a constant and performing a Fourier transformation, the above Hamiltonian writes as $H=\sum_{\bf k}\psi^{\dagger}_{\bf k}{\cal H}({\bf k})\psi_{\bf k}$, where $\psi_{\bf k}=\{ a_{A, {\bf k}}, a^{\dagger}_{B, -{\bf k}} \}^{T}$ is the basis, and
\begin{eqnarray}\label{eq8}
{\cal H}({\bf k})=\left[
                    \begin{array}{cc}
                      \Delta+\mu & f({\bf k}) \\
                      f^*({\bf k}) & \Delta-\mu \\
                    \end{array}
                  \right]
\end{eqnarray}
with $f({\bf k})=1+e^{-i{\bf k}\cdot{\bf a}_1}+e^{-i{\bf k}\cdot{\bf a}_2}$ [${\bf a}_1=(\sqrt{3},0),{\bf a}_2=(\sqrt{3}/2,3/2)$ the primitive vectors]. The above Hamiltonian should be diagonalized using Bogoliubov transformation ${\cal U}({\bf k})^{\dagger} {\cal H}({\bf k}){\cal U}({\bf k})=D$, where $D$ is a diagonal matrix containing the spectrum and ${\cal U}({\bf k})$ represents the Bogoliubov transformation. Due to the commutation relation of bosons ${\cal U}({\bf k})^{\dagger}s_z{\cal U}({\bf k})=s_z$, we have $s_z {\cal H}({\bf k}){\cal U}({\bf k})={\cal U}({\bf k})s_zD$. Thus to obtain the magnon spectrum, the following non-Hermitian matrix can be considered,
\begin{eqnarray}\label{eq9}
\sigma_z{\cal H}({\bf k})=\left[
                    \begin{array}{cc}
                      \Delta+\mu & f({\bf k}) \\
                      -f^*({\bf k}) & -(\Delta-\mu) \\
                    \end{array}
                  \right].
\end{eqnarray}
The eigenvalues are given by $E^{\pm}_{\bf k}=\mu \pm \epsilon({\bf k})$ with $\epsilon({\bf k})=\sqrt{\Delta^2-|f({\bf k})|^2}$.
The matrix of the eigenvectors is
\begin{eqnarray}\label{eq10}
{\cal U}_{\bf k}=\left[
                    \begin{array}{cc}
                     \cosh \theta_{\bf k}e^{i\phi_{\bf k}} & -\sinh \theta_{\bf k} \\
                      -\sinh \theta_{\bf k} & \cosh \theta_{\bf k}e^{-i\phi_{\bf k}} \\
                    \end{array}
                  \right],
\end{eqnarray}
where $\sinh 2\theta_{\bf k}=\frac{|f({\bf k})|}{\epsilon({\bf k})}$, $\tan \phi_{\bf k}=\frac{Im f({\bf k})}{Re f({\bf k})}$. The first (second) column is the eigenvector $u_{+,{\bf k}}$ ($u_{-,{\bf k}}$) corresponding to $E^{+}_{\bf k}$ ($E^{-}_{\bf k}$). The Hamiltonian in Eq.(8) is thus diagonalized by the transformation: ${\cal U}^{\dagger}_{\bf k}{\cal H}({\bf k}){\cal U}_{\bf k}=\textrm{diag}(E^{+}_{\bf k}, -E^{-}_{\bf k})$. The magnon spectrum consists of two branches, i.e., $E^{+}_{\bf k}, -E^{-}_{\bf k}$, which are plotted in Fig.\ref{fig5}(c).

The antiferromagnetic (AF) order of the spin model corresponds to the CDW insulator in terms of hardcore bosons. Thus the low-energy magnon bands, i.e, the excitation spectrum about the AF order, are related to the appearance of the superfluid right above the CDW insulator. When the spectrum becomes gapless, i.e., $E^{\pm}_{\bf k}=0$, superfluid begins to replace the CDW phase. Thus the condition $E^{\pm}_{\bf k}=0$ determines the phase boundary between the CDW and superfluid phases, which is $\mu=\pm \sqrt{\Delta^2-(3t)^2}$. We plot the phase boundary from the spin-wave approximation in Fig.\ref{fig5}(a). It is qualitatively consistent with the exact phase diagram from the QMC method except that the region is reduced in the $(\Delta/t, \mu/t)$ phase. We also determine the phase boundary of the $\rho=\frac{1}{2}$ domain-wall phase on a $W=12$ zigzag ribbon, and the upper boundary is slightly shifted downward, which is also consistent with the QMC result.

The Berry curvature associated with each magnon band is given by~\cite{berry1984,zhang2016}
\begin{eqnarray}\label{eq11}
\Omega_{\lambda}({\bf k})=\frac{\partial A_{y}({\bf k})}{\partial k_x}-\frac{\partial A_{x}({\bf k})}{\partial k_y},
\end{eqnarray}
where $A_{i}=-i\langle u_{\lambda,{\bf k}}|\frac{\partial}{\partial k_i}|u_{\lambda,{\bf k}}\rangle$ ($i=x,y$) is the Berry potential, and $\lambda=\pm$ denotes the two magnon bands\cite{fukui2005}.

As shown in Fig.\ref{fig5}(d), the Berry curvature is peaked at the corners of the Brillouin zone (BZ), and is antisymmetric with respect to the inversion center ${\bf k}=(0,0)$. The Berry curvatures for the two $\rho=\frac{1}{2}$ CDW insulators differ by an overall sign. Although the sum of the Berry curvature of each band in the BZ (known as the Chern number) vanishes identically, there is a sign change for the Berry curvature across the domain wall, which may results in gapless boundary phase\cite{semenoff2008}. In the spin-wave approximation, the magnon spectrum of a zigzag ribbon has two such branches associated with the domain wall [see Fig.\ref{fig5}(b)]. One of them is at the bottom of the spectrum, and it corresponds to the superfluid above the $\rho=\frac{1}{2}$ CDW insultor, which is localized near the domain wall.

\begin{figure}[htbp]
\centering \includegraphics[width=7.cm]{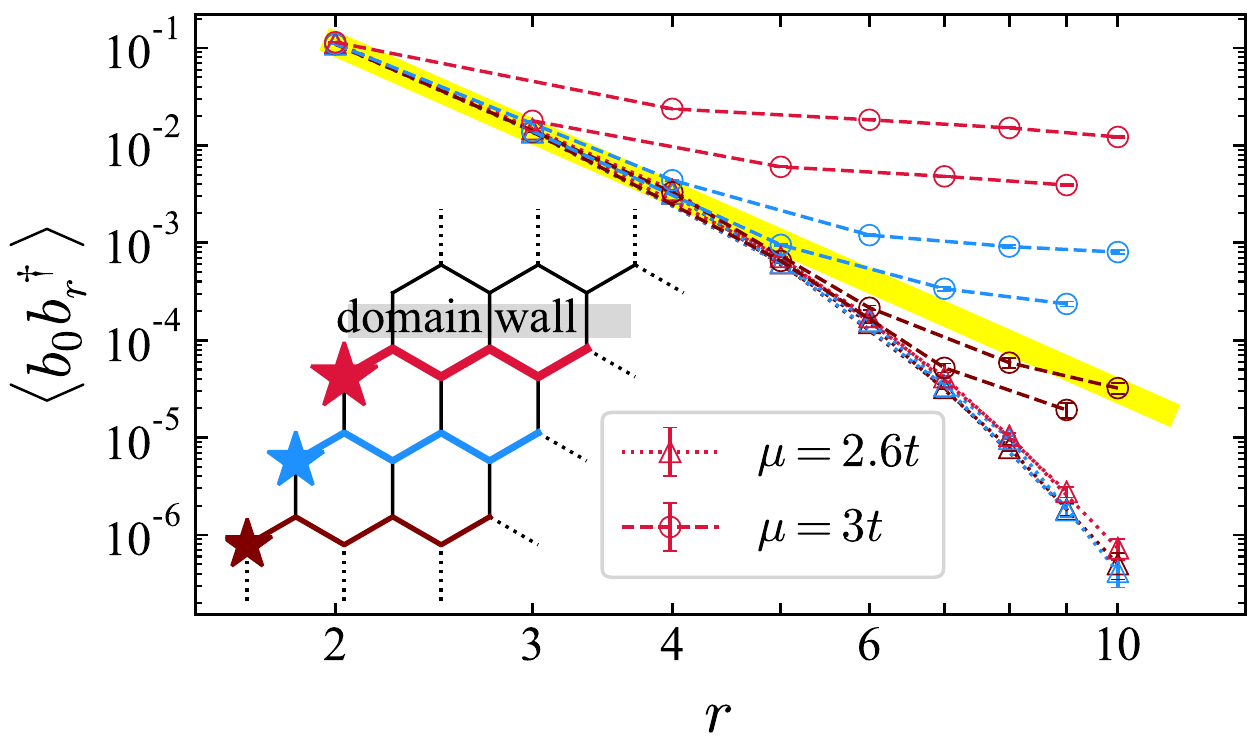}
\caption{The single-particle correlator
   $\langle b^{\dagger}_{0} b_{r}\rangle$ along several nonequivalent zigzag chains near the domain wall. The star symbols on the
inset geometries mark the reference site $r=0$.
The thick yellow lines are plotted as guides to algebraic behavior. Up triangles connected by dotted lines refer to the insulating regime with $\rho = \frac{1}{2}$. Ribbon width $W=12$ and length $L=24$.
}
\label{fig6}
\end{figure}

To verify the localization of the superfluid near the domain
wall, we calculate the single-particle correlator $\langle
b^{\dagger}_{0}b^{\phantom{\dagger}}_{r}\rangle$
using QMC, as shown in Fig.~\ref{fig6}. The correlator along the
zigzag chain on the domain wall
is slower than a power-law
decay with distance, which is characteristic of a gapless quasi-1D superfluid. In
contrast, the excitation is gapped for the $\rho=\frac{1}{2}$
domain-wall insulator, and the correlator decays exponentially. As
one moves away from the domain wall, the correlator becomes increasingly
short-ranged, and $\rho_s$ decreases. We have checked the superfluid
density decays exponentially with the distance away from the domain
wall.

\section{The armchair domain wall}

\begin{figure}[htbp]
\centering \includegraphics[width=9.cm]{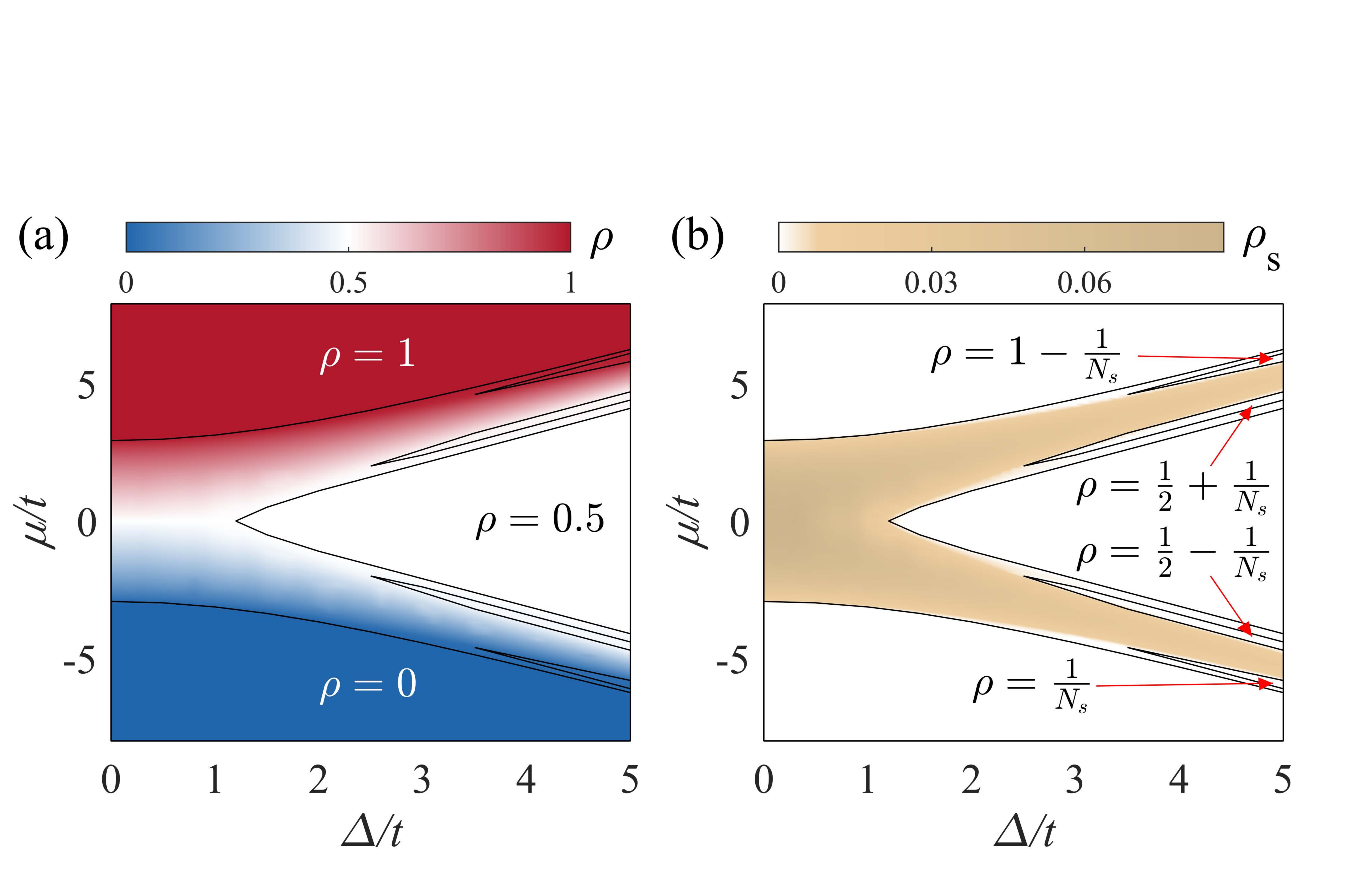} \caption{The phase diagram of a honeycomb lattice ribbon with an armchair domain wall in the $(\Delta, \mu)$ plane, which contains superfluid and insulators at various fillings. The false color plots of the average density and the superfluid density are shown in (a) and (b), respectively.}
\label{fig7}
\end{figure}

\begin{figure}[htbp]
\centering \includegraphics[width=7.5cm]{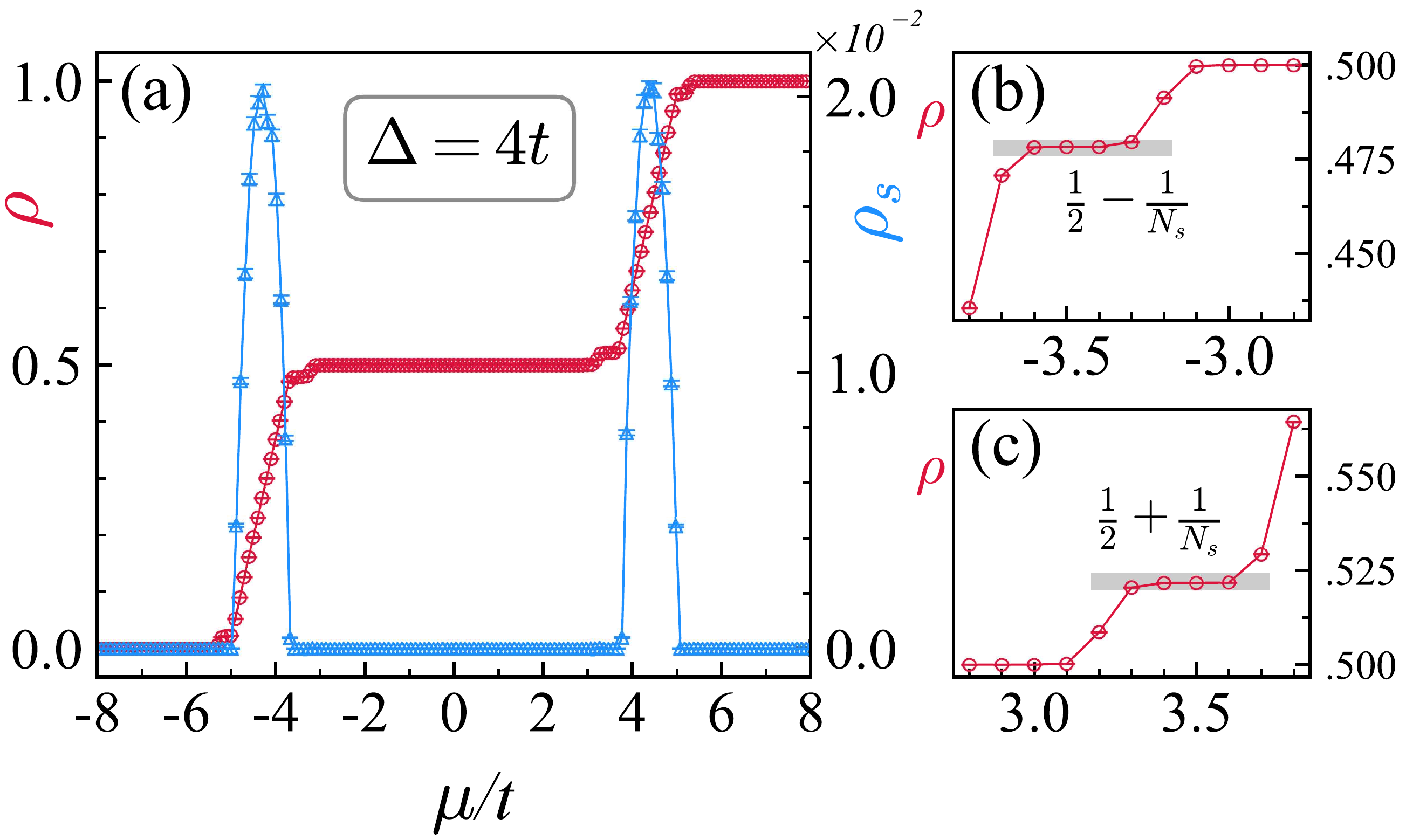} \caption{(a) The average density and superfluid density as a function of $\mu$ at $\Delta=4t$ on a $W=12$ and $L=24$ ribbon with an armchair domain wall in the middle. (b) and (c) are enlarged illustrations of the two small insulating regions near $\rho=\frac{1}{2}$. Here the number of sites in each unit cell is $N_s=4*W-2$, and thus the values of the plateaus in (b) and (c) are $\rho=0.4783$ and $0.5217$, respectively.}
\label{fig8}
\end{figure}

Next we study the Bose-Hubbard model on a honeycomb ribbon with an armchair domain wall in the middle. Figure \ref{fig7} shows the phase diagram in the $(\Delta,\mu)$ plane, along with the false color plots of the average density [see Fig.\ref{fig7}(a)] and the superfluid density [see Fig.\ref{fig7}(b)]. In the atomic limit, the system is in the $\rho=0$ empty phase for $\mu<\Delta$, the $\rho=1$ Mott insulator for $\mu>\Delta$, and the $\rho=\frac{1}{2}$ domain-wall insulator for $-\Delta<\mu<\Delta$. Besides the three atomic like phases, there appear four small insulating regions at large $\Delta$, which are symmetric about $\mu=0$. To see the details of the various phases, we plot $\rho$ and $\rho_s$ as a function of $\mu$ along a typical cut $\Delta= 4t$ in Fig.\ref{fig8}. The small insulating regions are located at both ends of the transition region between the $\rho=0$ and $\frac{1}{2} $(also $\rho=\frac{1}{2}$ and $1$) insulators. Their average densities are $\rho=\frac{1}{N_s},\frac{1}{2}-\frac{1}{N_s}, \frac{1}{2}+\frac{1}{N_s}$, and $1-\frac{1}{N_s}$, respectively, where $N_s$ is the number of sites in each unit cell . For a bond crossed by the domain wall, both sites connected have the same on-site potentials. Each unit cell contains one pair of low-potential sites and one pair of high-potential sites, and the above insulators are closely related to the occupying statuses of such sites.

For large $\Delta$, the pairs of low(high)-potential sites on the domain wall form dimers. The gain of kinetic energy for one boson in each dimer is approximately $-t$, while that in the CDW phase is proportional to $-\frac{t^2}{2\Delta}$, which is due to the second-order process and much smaller then $-t$. Starting from the empty phase, the bosons first enter the low-potential sites on the domain wall due to the large gain of kinetic energy, and the resulting insulator consists of isolated dimers. Since each unit cell has one such dimer, the average density is $\frac{1}{N_s}$. As the chemical potential increases, the low-potential sites away from the domain wall become occupied, and the system is a $\rho=\frac{1}{2}-\frac{1}{N_s}$ insulator. Then the empty low-potential sites on the domain wall are occupied, and it is a $\rho=\frac{1}{2}$ insulator. When the chemical potential is large enough, the bosons begin to occupy the high-potential sites. Once again, they first enter such sites on the domain wall, forming a $\rho=\frac{1}{2}+\frac{1}{N_s}$ insulator. To maintain the large gain of the kinetic energy, the bosons next occupy the high-potential sites away from the domain wall resulting in a $\rho=1-\frac{1}{N_s}$ insulator. Finally the empty sites on the domain wall are occupied, and the system becomes full.

Although the bosons do not simply fill into the bands like the fermions, the feature of the band structure [see Fig.\ref{fig1}(d)] is reflected. As has been stated, there are four bands associated with the armchair domain wall, two of which are inside the gap and two are outside the bulk bands. The small insulating regions correspond to the gaps between the domain-wall bands and the bulk ones. Due to the condensing nature of the bosons, the gap sizes are reduced, and the gaps persist only for large $\Delta$. Thus such regions appear only at large $\Delta$, which is evident in Fig.\ref{fig7}.

There also appears domain-wall superfluid between each of the small insulating regions and the adjacent commensurate insulator. However the superfluid density is much smaller than that associated with a zigzag domain wall. The reason is that an armchair domain wall consists of alternating two high-potential sites and two low-potential ones, thus the superfluid transport down such a domain wall is greatly reduced.

\begin{figure}[htbp]
\centering \includegraphics[width=9.cm]{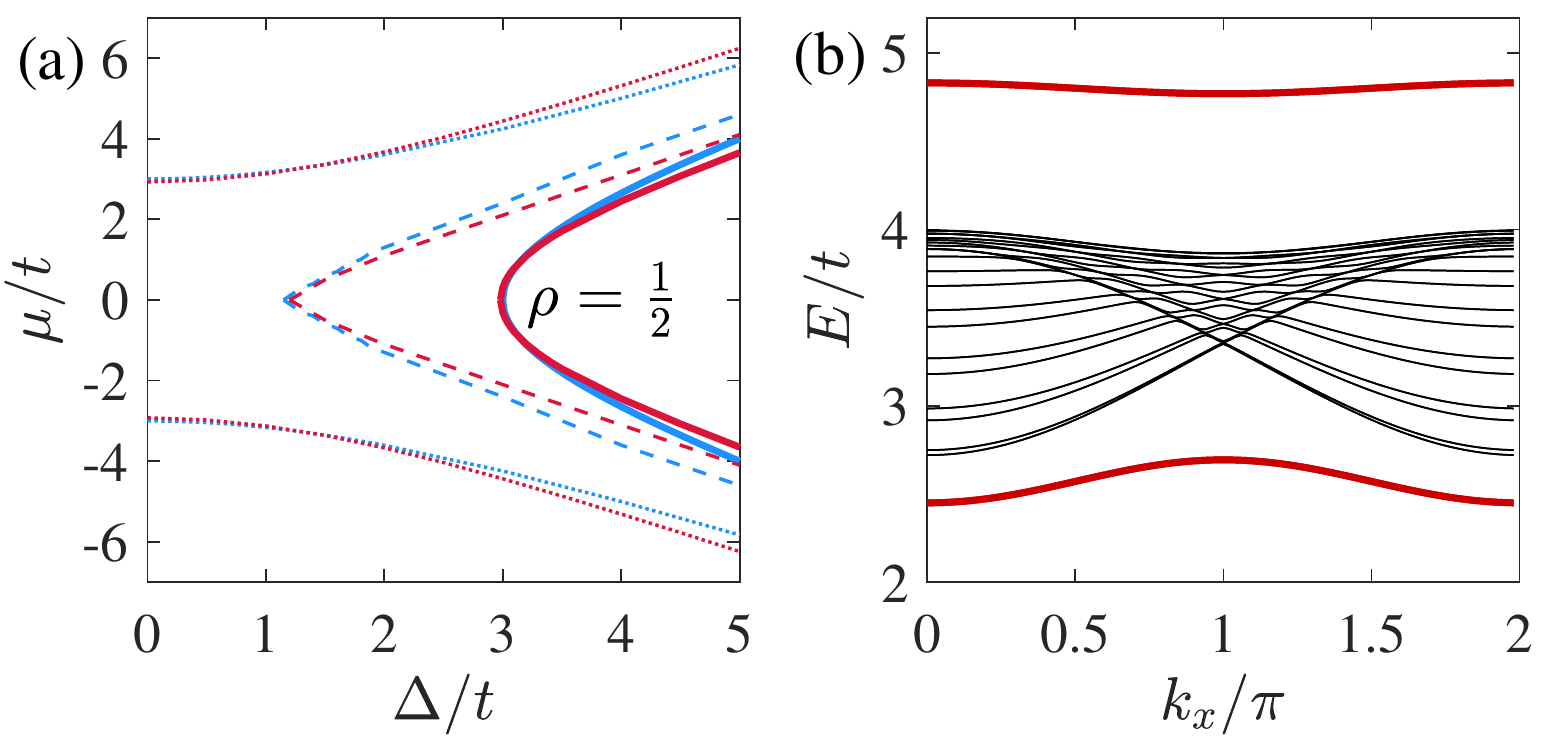}
\caption{(a) The phase boundaries from the spin-wave approximation: the thick solid blue curve is for the periodic honeycomb lattice with a staggered potential; the thick solid red curve is for the armchair ribbon with a domain wall in the middle. We also plot exact QMC results for comparison (dashed and dotted color lines). (b) The excitation spectrum on a $W=12$ ribbon with an armchair domain wall in the middle. The red curves represent states localized near the armchair domain wall. The parameters are $\Delta=4t$.
}
\label{fig9}
\end{figure}

We also performed a mean-field analysis for the armchair domain wall. The phase boundary of $\rho=\frac{1}{2}$ insulator is shown in Fig.\ref{fig9}(a). Similar to the zigzag case, the size of the $\rho=\frac{1}{2}$ region is shrunk in the linear spin-wave approximation. Due to the particle-hole symmetry, the phase boundary is symmetric about $\mu=0$, which is contrast to the asymmetric zigzag case. Corresponding to the nontrivial bulk topological property of Bose-Hubbard model with a staggered potential, there appear isolated domain-wall bands in the magnon spectrum on the armchair ribbon with a domain wall in the middle, as shown in Fig.\ref{fig9}(b).

\section{Conclusions}
We study hardcore bosonic domain walls on honeycomb lattice gapped by a staggered potential using QMC simulations. The phase diagrams contain the superfluid and insulator phases at various fillings. It is revealed that the $\rho=\frac{1}{2}$ insulator is a domain wall phase, where the domain wall separates two CDW regions with opposite Berry curvatures. Associated with this superfluid transport occurs down the domain wall. Due to the different arrangements of occupied and unoccupied sites along the domain wall, the superfluid density associated with a zigzag domain wall is much larger than that of an armchair domain wall.
Experimentally the Bose-Hubbard model can be simulated using cold atoms trapped in optical lattices, and staggered on-site potentials are naturally realized\cite{sebby2006,esslinger2015}. The honeycomb geometry has been readily obtained with three laser beams intersecting at an angle of 120 degree ~\cite{Polini2013, zhu2007, soltanpanahi2011}. New observation tools based on quantum gas microscopes allow observation of the density profile at the level of individual
atoms\cite{Gross2017,bloch2012,bakr2009,sherson2010,chin2009}. Besides, the Berry curvature can be directly measured via interferometric techniques\cite{Duca2015,Flaschner2016}. With these state-of-art technologies, it is very possible that our results are demonstrated experimentally.

\section{Acknowledgments}
The authors thank Prof. R. Scalettar and Prof. R. Mondaini for helpful discussions. H.G. acknowledges support from the NSFC grant No.~11774019.
X.Z. and S.F. are supported by the National Key Research and Development Program of China under Grant No. 2016YFA0300304, and NSFC under Grant Nos. 11974051 and 11734002

\appendix
\section{The eigenenergies of the zigzag domain wall at $k_x=\pi$}

The energy spectrum of the zigzag domain wall can be analytically derived at $k_x=\pi$. Choosing the unit cell shown in Fig.\ref{fig1}(a), the Hamiltonian in the momentum space writes as,
\begin{eqnarray}\label{aeq1}
H_{z}(k_x)=\left(
                      \begin{array}{cccccc}
                        -\Delta & h(k_x) & 0 & 0 & 0 & ... \\
                        h^*(k_x) & \Delta & -t & 0 & 0 & ... \\
                        0 & -t & -\Delta & h^*(k_x) & 0 & ... \\
                        0 & 0 & h(k_x) & \Delta & -t & ... \\
                        0 & 0 & 0 & -t & \Delta & ... \\
                        ... & ... & ... & ... & ... & ... \\
                      \end{array}
                    \right),
\end{eqnarray}
with $h(k_x)=-t(1+e^{-ik_x})$. The momentum $k_x=\pi$ is special, where $h(\pi)=0$. The Hamiltonian matrix becomes block diagonal, containing a series of $2\times 2$ matrices describing the localized states in each vertical bond, and two isolated elements representing the outmost sites. There are two kinds of vertical bonds. While the one on the domain wall has uniform on-site potentials which has been discussed in the main text, the other one has opposite potentials on the two sites connected by the vertical bond, and the matrix is
\begin{eqnarray}\label{aeq2}
H_{k_x=\pi}=\left(
              \begin{array}{cc}
                -\Delta & -t \\
                -t & \Delta \\
              \end{array}
            \right).
\end{eqnarray}
The eigenvalues are $\pm\sqrt{t^2+\Delta^2}$. Since the number of the bonds described by the above matrix increases with the width of the ribbon, such localized states are multifold degenerate, which can be seen in Fig.\ref{fig1}(c).

The top-left and bottom-right blocks of Eq.(\ref{aeq1}) containing a single element describe localized states on the outmost sites, whose eigenenergy is simply the on-site potential $-\Delta$.

% The \nocite command causes all entries in a bibliography to be printed out
% whether or not they are actually referenced in the text. This is appropriate
% for the sample file to show the different styles of references, but authors
% most likely will not want to use it.
%\nocite{*}
\nocite{*}
\bibliography{domain_ref}% Produces the bibliography via BibTeX.

\begin{thebibliography}{50}
\bibitem [{bul()}]{bulk_bands}%
  \BibitemOpen
  \href@noop {} {}\bibinfo {note} {Here the bulk bands refer to the extended
  bands in the spectrum of the ribbon, which are reminiscent of those in a bulk
  system with periodic boundary conditions in both directions.}\BibitemShut
  {Stop}%
\end{thebibliography}

\appendix

\end{document}